\documentclass{aa}
\usepackage{graphicx}
\usepackage{natbib}
\usepackage{amsmath}
\bibpunct{(}{)}{;}{a}{}{,}

\newcommand{\PS}{{\rm PS}}
\newcommand{\IPS}{{\rm IPS}}
\newcommand{\FS}{{\rm FS}}
\newcommand{\DFS}{{\rm DFS}}

\newcommand{\WVD}{{\rm WVD}}
\newcommand{\GS}{{\rm GS}_D}
\newcommand{\FSR}{{\rm FS}_r}
\newcommand{\FT}{{\rm FT}}
\newcommand{\STFT}{{\rm STFT}}
\newcommand{\DSTFT}{{\rm DSTFT}}
\newcommand{\SDFS}{{\rm SDFS}}
\newcommand{\SCL}{{\rm SCL}}
\newcommand{\WT}{{\rm WT}}
\newcommand{\w}{{\bf w}^*}

\newcommand{\ACF}{{\rm ACF}}

\begin{document}

\title{Joint Time--Frequency Analysis: \\
a tool for exploratory analysis and filtering of \\ non-stationary time series}

   \author{R. Vio\inst{1}
          \and
          W. Wamsteker\inst{2}
          }

   \offprints{R. Vio}

   \institute{Chip Computers Consulting s.r.l., Viale Don L.~Sturzo 82, \\
              S.Liberale di Marcon, 30020 Venice, Italy\\
             ESA-VILSPA, Apartado 50727, 28080 Madrid, Spain\\
              \email{robertovio@tin.it}
         \and
             ESA-VILSPA, Apartado 50727, 28080 Madrid, Spain\\
             \email{willem.wamsteker@esa.int}
             }

\date{Received 24/07/01; accepted 03/04/02}

\abstract{
It is the purpose of the paper to describe the virtues of time-frequency methods
for signal processing applications, having astronomical time series in mind.
Different methods are considered and their potential usefulness respectively
drawbacks are discussed and illustrated by examples. As areas where one can
hope for a successful application of joint time-frequency analysis (JTFA), we describe 
specifically the problem of signal denoising as well as the question of signal separation 
which allows to separate signals (possibly overlapping in time or frequency, but) which
are living on disjoint parts of the time-frequency plane. Some
recipes for practical use of the algorithms are also provided.
}
\titlerunning{JTFA of Non-Stationary Time Series}
\authorrunning{R. Vio \& W. Wamsteker}
\maketitle

\section{Introduction}

The analysis of time series has always been an issue of central interest to
astronomers. Indeed, the most natural
way to get some insights on the characteristics of a physical system consists in
studying its behavior over time. Unfortunately, maybe because until  few years ago
the most interesting objects known to change their luminosity were periodic/semiperiodic
stars and multiple star systems, time series analysis became
synonymous of research of periodicities and power-spectrum became the tool for
signal analysis. Only in the eighties people began to realize that various
astrophysical objects, principally extragalactic sources, can show very complex
and unpredictable time evolutions \citep[see, for example,][]{mil91, dus91}. In spite of that, 
although inadequate, power-spectrum continued to be the preferred approach by most of 
researchers.
Such a predilection can be explained with the substantial failure of the methods
proposed as alternatives \citep{vio92}. The main limitations of such methods are two:
\begin{enumerate}
\item[-] The implicit assumption of stationarity of the processes underlying the signals. 
In other words, an observed signal is assumed to be characterized by statistical properties 
that do not change with time. However, real signals in their vast majority are nonstationary.
\item[-] The implicit assumption that it is possible to obtain the equations describing the 
evolution of a given system solely through the analysis of an experimental signal. In reality, 
if one takes into account that a given time series is only the one-dimensional output of a 
system which very often can be described only in terms of a set of ordinary and/or partial 
differential
equations (in other words a time series is the one-dimensional projection of a
more complex dynamics), it is easy to  understand that this position is not well founded.
\end{enumerate}
The consequence of these points is that if not carried out in a well defined physical
context, signal analysis can be expected to be useful only for explorative
purposes (e.g. feature detection) and/or for filtering tasks (e.g. removal of
unwanted components as, for example, noise or spurious contributions). Furtehrmore, in order 
to obtain meaningful results, the nonstationarity of signals must be explicitly considered in 
the analysis.

\section{Data exploration}
\subsection{Limits of the classical approach}

It is well known that the power-spectrum $\PS(\nu)$ of a continuous and integrable
(possibly complex) signal $x(t)$ is defined as
\begin{equation} \label{eq:PS}
\PS(\nu)= | \FT(\nu) |^2,
\end{equation}
where the frequency range is the full real line, i.e. \, $-\infty < \nu < +\infty$, and
\begin{equation} \label{eq:FT}
\FT(\nu)=\int\limits_{- \infty}^{+ \infty} x(t)~e^{-j 2 \pi \nu t}~dt
\end{equation}
is the Fourier transform of $x(t)$. $\PS(\nu)$ can be considered as a measure of
the similarity between signal $x(t)$ and the complex sinusoid $\exp(- j 2 \pi
\nu t)$. Indeed, if $\FT(\nu)$ is again integrable, the Fourier inversion formula reads
\begin{equation}
x(t) =\int\limits_{- \infty}^{+ \infty} \FT(\nu)~e^{j 2 \pi \nu t}~d\nu,
\end{equation}
i.e., the signal can be written (uniquely) as a kind of superposition
of pure frequencies.
Therefore, the larger $\PS(\nu)$, the more important is the contribution
of the pure frequency $\nu$ within $x(t)$.
\begin{figure}
	\resizebox{\hsize}{!}{\includegraphics{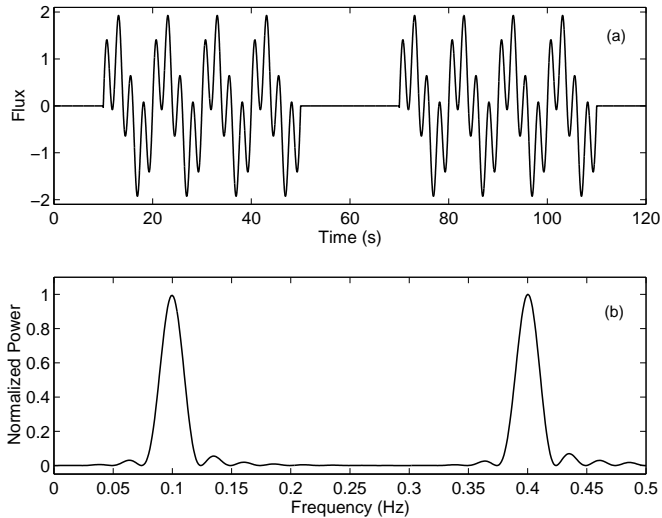}}
	\caption{ {\bf a)} Realization of the non-stationary
	process described in the text. {\bf b)} Corresponding normalized
	power-spectrum.}
	\label{fig:sinus2}
\end{figure} 
This is the reason why $\PS(\nu)$ is so useful in the
search of periodic components. However, this is also its more serious limit in
the analysis of non-periodic signals \citep{fei98}. Fig.\ref{fig:sinus2}a shows a signal
constituted by two sinusoids with amplitudes different from zero only within a
short time interval:
\begin{equation}
x(t)=a_1(t) \sin(2 \pi \nu_1 t) + a_2(t) \sin(2 \pi \nu_2 t)
\end{equation}
with $\nu_1=0.1~\mbox{Hz}$, $\nu_2=0.4~\mbox{Hz}$ and
\begin{equation}
a_1(t), a_2(t) = \left\{
\begin{array}{ll}
1 & \mbox{if $ 10 \leq t \leq 50$ and $70 \leq t \leq 110$
}; \\ 0 & \mbox{otherwise},
\end{array}
\right.
\end{equation}
with $t$ expressed in seconds.
The two key facts are:
\begin{itemize}
\item[-] this signal clearly is not periodic and
\item[-] it does not maintain its characteristics constant over time
         (in poor words, it is non-stationary).
\end{itemize}
From Fig.\ref{fig:sinus2}b it is evident that $\PS(\nu)$ is not able to extract
the salient features of $x(t)$.
In particular, the power spectrum clearly contains no information about the time
intervals on which these two frequencies occur. Moreover, since $\PS(\nu) \neq 0$ also at
frequencies $\nu \neq \nu_1,\nu_2$, one could erroneously infer that $x(t)$
contains more than two sinusoidal components. The main problem is that, if a
signal changes its characteristic over time, a representation limited exclusively
to the frequency domain is absolutely inadequate. Indeed, the function
$\exp(- j 2 \pi \nu t)$ does not have any temporal localization: it has neither start nor
end. This is an useful characteristic only in case of signals with constant
statistical properties (i.e. periodic or stationary signals).
Unfortunately, in nature this kind of signals is not so common.
Therefore, for a good analysis of most ``natural'' signals it is necessary to use a
representation that is able to consider also the temporal dimension.
A possible solution is a representation, that we call {\it ideal power-spectrum}, 
$\IPS(t,\nu)$, defined as the squared amplitude of the sinusoids that
at a given time instant $t$ constitute the signal of interest.
\begin{figure}
	\resizebox{\hsize}{!}{\includegraphics{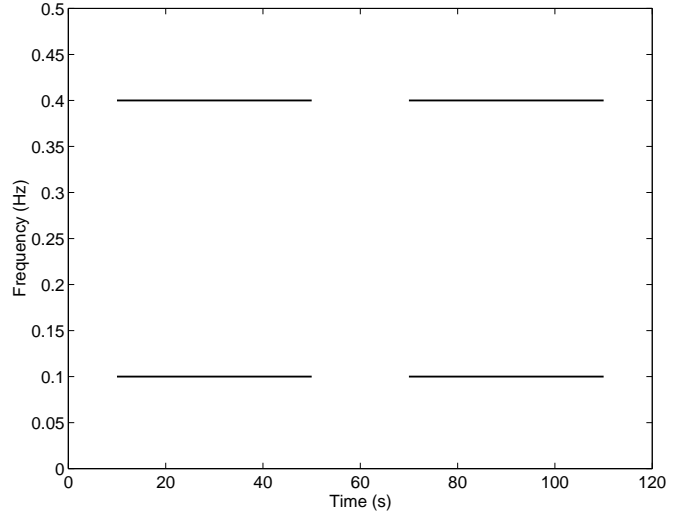}}
	\caption{ Grayscale image of the ideal power-spectrum $\IPS(t,\nu)$ 
	corresponding to the signal shown in Fig.\ref{fig:sinus2}a.}
	\label{fig:tf_sinus2}
\end{figure} 
That is shown in Fig.\ref{fig:tf_sinus2} where it is possible to realize that with 
$\IPS(t,\nu)$, contrary to $\PS(\nu)$, the characteristics of the signal in 
Fig.\ref{fig:sinus2}a are very well defined; it is clear that such a signal consists 
of two pure sinusoids having different frequencies but amplitudes which evolve in the same 
way.

\subsection{Time--frequency representation of continuous signals}

In the previous section we have seen that in case of non-stationary signals a joint
time-frequency representation (JTFR) might be a good analysis tool.
However, the JTFR in Fig.\ref{fig:tf_sinus2} has been obtained directly
from the model with which the time series has been simulated.
Unfortunately, in an experimental context the dynamical model
underlying a given time series will usually be unknown.
Consequently, the problem consists in the way such a JTFR can
be estimated from a set of available data. The simplest solution is
given by the so called {\it Fourier Spectrogram}, $\FS(t,\nu)$,
\begin{equation} \label{eq:Loc_PS}
\FS(t,\nu)= | \STFT(t,\nu) |^2,
\end{equation}
where $-\infty < t < + \infty, ~-\infty < \nu < + \infty$, and
\begin{equation} \label{eq:Loc_FT}
\STFT(t,\nu)=\int\limits_{- \infty}^{+ \infty} w^*(t-\tau)~x(\tau)~e^{-j 2 \pi \nu
\tau}~d\tau,
\end{equation}
(symbol ``${}^*$'' stands for complex conjugation), is the so called {\it short time fourier 
transform}. Here
\begin{equation}
\begin{array}{lll}
w(t) & \neq 0 & \qquad \mbox{for $ -T/2 \leq t \leq T/2$}; \\
 & \simeq 0 & \qquad \mbox{otherwise} \\
\end{array}
\end{equation}
is a window function usually indicated with the term of {\it analysis function}.
The rationale is that, for signals  $x(t)$ which maintain
their harmonic components unchanged at least approximately
within a time interval of length $T$ and centered at $t_o$,
$\FS(t_o,\nu)$ can be considered an estimate of $\IPS(t_0,\nu)$. Although
effective, this methods is characterized by two drawbacks:
\begin{itemize}
\item[-] since $\FS(t,\nu)$ is estimated via the Fourier transform of only a
segment of the available signal, it suffers a degradation of the frequency
resolution {\it inversely} proportional to the length of $w(t)$.
For example, if $w(t)$ is the rectangular window
\begin{equation}
w(t) = \left\{
\begin{array}{ll}
1 & \mbox{for $ -T/2 \leq t \leq T/2$}; \\ 0 & \mbox{otherwise}, \\
\end{array}
\right.
\end{equation}
it is easy to show that
\begin{equation}
\STFT(t_o,\nu) = \FT(\nu) \otimes {\rm sinc}_T(t_o,\nu),
\end{equation}
where the symbol ``$\otimes$'' stands for convolution, and
\begin{equation}
{\rm sinc}_T(t_o,\nu)=\frac{\sin(T \pi \nu)}{\pi \nu}~e^{- j 2 \pi \nu t_o}.
\end{equation}
This means that, in $\FS(t_o,\nu)$, the narrowest structure along the frequency direction is 
charaterized by a frequency support equal to $2/T$. Such value corresponds to the {\it full 
width at zero intensity} of the main lobe in the graphical representation of $|{\rm
sinc}_T(t_o,\nu)|^2$;
\item[-] since $\FS(t,\nu)$ is calculated by using all the points in a interval
of length $T$ centered at $t$, it suffers also a degradation of the resolution along
the time dimension which is {\it proportional} to the length of $w(t)$.
Indeed, in case of a signal constituted by an impulse at time $t_0$,
and again with a rectangular $w(t)$, $\FS(t,\nu) \neq 0$
when $t_o - T / 2 \leq t \leq t_o + T / 2$. In other words, the narrowest structure along the 
time direction is charaterized by a time support equal to $T$. 
\end{itemize}
From these two points it is easy to understand that the main limitation of
$\FS(t,\nu)$ lies in the impossibility to increase simultaneously the time
and the frequency resolutions. In other words, there is a trade-off: if
$w(t)$ is chosen to have good time resolution (i.e., smaller $T$), then its
frequency resolution must be deteriorated and vice versa. In principle, a
possible way to mitigate this occurrence is the choice of a window $w(t)$
for which the degradation of the time-frequency resolution is reduce to a
minimum. In this regard, it has been proved \citep[e.g.,][]{qia96} 
that, if the time-frequency resolution of a representation is measured by its energy 
concentration in the time-frequency domain, then the {\it Gaussian} function
\begin{equation}
w(t)=\left( \frac{\alpha}{\pi} \right) ^{\frac{1}{4}} e^{-\frac{\alpha}{2} t^2}
\end{equation}
represents an ``optimal'' choice. Without any a priori information on the spectral 
characteristics of the signal, $\alpha$ is a free parameter with which it is possible to 
balance between
an increased time or frequency resolution according to the current necessities.
In reality, if $w(t)$ is a smooth function, well localized around zero,
it will serve well as a window for the above purpose,
independently from the precise analytic form.
In this regard note that, because of its lack of smoothness, the rectangular window, used
above only for ease of discussion, does not represent a good choice because
of its very poor frequency resolution.

\subsection{Time-frequency representation of discrete signals}

In practical situations a signal is available only in discrete form, say $x[k]$,
obtained by sampling $x(t)$ on a regular grid of $L_s$ time instants
$k = 0, 1, 2, \ldots, L_s-1$. Consequently, $\FS(t,\nu)$ and $\STFT(t,\nu)$ have to be
modified to their discrete variants:
\begin{eqnarray}
\DFS[m,n] &=& | \DSTFT[m,n] |^2; \label{eq:Loc_DPS} \\ \DSTFT[m,n] &=&
\sum_{k=0}^{L_s-1} w^*[k-m]~x[k]~e^{- j 2 \pi n k / L_s}, \label{eq:Loc_DFT}
\end{eqnarray}
with $m$, $n$, $0 \leq m, n \leq L_s-1$, representing the discrete time and
frequency indices, respectively.

\begin{figure}
	\resizebox{\hsize}{!}{\includegraphics{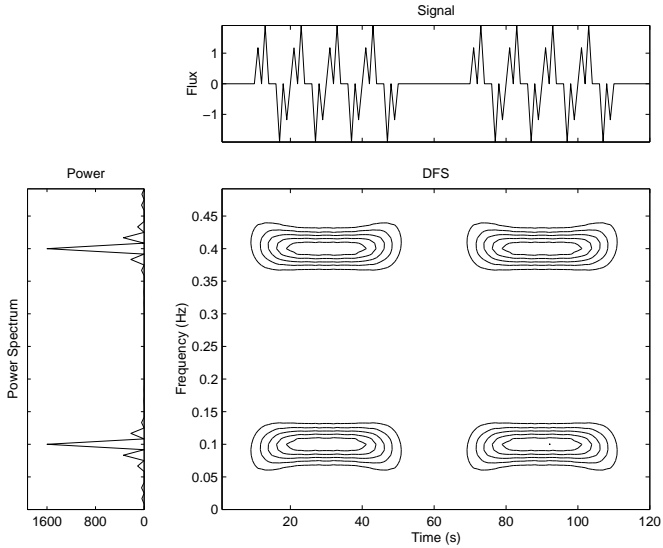}}
	\caption{ Contour plot of the $\DFS[m,n]$ representation
	for the discrete time series, shown in the uppermost panel, corresponding to the 
	signal in 	Fig.\ref{fig:sinus2}a when sampled at a frequency rate of $1.0$ Hz 
	(the corresponding power-spectrum 	is shown in the leftmost panel). 
	In this example the analysis function is an
	{\it Hamming} window with length $L_w = 31$. Levels are uniformed distributed
	between the $15\%$ and $85\%$ of the peak value.}
	\label{fig:tf_dsinus2}
\end{figure}

\noindent Fig.\ref{fig:tf_dsinus2} shows the representation $\DFS[m,n]$ of a time
series obtained by sampling the signal in Fig.\ref{fig:sinus2} with a time step
of one second: in spite the loss of resolution described above, the salient
characteristics of the signal remain well defined.

\subsection{Some practicalities}

\subsubsection{Digital implementation}

The calculation of $\DFS[m,n]$ requires some technical adjustments. The first one
regards the edge effects, due to the finite length of the signal, that do not
permit the calculation of this representation for $m < L_w/2$ and $m >
L_s-1-L_w/2$, with $L_w$ the length of $w[k]$. Usually this problem is solved
through one of the following three methods:
\begin{itemize}
\item[1)] by assuming $x[k]$ and $w[k]$ to be periodic with period $L_s$
           (NB. since usually $L_w < L_s$, $w[k]$ has to be zero padded);
\item[2)] by zero padding one of the ends of $x[k]$ with $L_w$ data and
     assuming the new time series to be periodic with period $L_s + L_w$
     \citep[for an efficient implementation of this method see][]{qia96};
\item[3)] by imposing that $0 \leq m \leq M_{T}$, with $M_{T} = L_s - L_w + 1$
        the number of possible shifts of $w[k]$ inside the signal length \citep[for an
       efficient implementation of this method see][]{mun97, mun01};
\end{itemize}
If not stated otherwise, in the following the discussion will be developed
according to the first approach.

A more serious drawback is represented by the fact that, by increasing the length
of $x[k]$, the corresponding $\DFS[m,n]$ becomes quickly an huge matrix. Indeed,
already $L_s = 1000$ will result in a matrix with $\simeq 10^6$ elements. A
viable solution is to cut the signal in shorter segments and then to analyze them
separately. An alternative comes out from the observation that $\DFS[m,n]$ is an
highly redundant representation; it is composed by $\simeq L_s^2$ elements that,
however, have been obtained from only $L_s$ data. Since $L_s$ is the dimension
of the signal space, at most $L_s$ elements can be linear independent.
This means that, in principle, it is possible to recover arbitrary signals if
samples of $\DFS[m,n]$ for at least $L_s$ coordinates are available. According to 
this observation, Eqs.(\ref{eq:Loc_DPS}),(\ref{eq:Loc_DFT}) could be modified in
\begin{equation}
\SDFS[m,n] = | a_{mn} |^2; \label{eq:Loc_DPS1}
\end{equation}
\begin{eqnarray} \label{eq:Loc_DFT1}
a_{mn} &=& \DSTFT[m \Delta M,n \Delta N ] \nonumber \\[-3mm] & &\\[-3mm] &=&
\sum_{k=0}^{L_s-1} w^*[k - m \Delta M]~x[k]~e^{- j 2 \pi n k \Delta N / L_s}
\nonumber
\end{eqnarray}
where $0 \leq m < M = L_s / \Delta M$, $0 \leq n < N = L_s/ \Delta N$, with
$\Delta M$ and $\Delta N$ integer divisors of $L_s$ such that $L_s/(\Delta M
\times \Delta N) \geq 1$.
In such a signal representation
the coefficients $a_{mn}$ are called the {\it Gabor coefficients} and
the equality (\ref{eq:Loc_DFT1}) goes by the name {\it Gabor representation} (GR).

Given the above arguments, in order to avoid ``unnecessary'' redundancy, one could be 
tempted to use a $\SDFS[m,n]$ containing only $L_s$ elements.
In reality, numerical stability of the expansion
and easiness of the analysis \citep[][ and see below]{dau92}, make it necessary
to work with redundant representations. The degree of redundancy is
expressed through the so called {\it oversampling rate}, say $\alpha$,
\begin{equation}
\alpha = \frac{\mbox{number of elements of $\SDFS[m,n]$}}{\mbox{length of
$x[k]$}},
\end{equation}
that will take values in the range $[1, L_s]$.
When $\alpha=1$, $\SDFS[m,n]$
is said to be sampled at a {\it critical rate}, whereas when $\alpha > 1$ it is said
{\it oversampled}. Using a full short-time Fourier transform corresponds to the case 
$\alpha = L_s$.

\subsubsection{Which $L_w$ for $w[k]$?}

One important step in the construction of JTFR's, such as $\SDFS[m,n]$, is the
choice of the analysis function $w[k]$.
Although the exact shape
of this function is not so critical, the value of $L_w$ can severely
influence the final result. In fact, in order to be meaningful, a $\SDFS[m,n]$
has to reflect both the instantaneous frequency content and the (possible)
non-stationarity of the signal. However, very often these two requirements
conflicts one with the other. The reason is that a large $L_w$ is preferable when
the frequency content is of interest and shows little changes,
whereas the time evolution of a signal can be better tracked with
a small $L_w$. Again, we are forced to decide within a  trade-off.

Despite its importance and its long-time use in signal processing,
surprisingly this subject has found little attention
in current literature \citep[however, see ][ and references therein]{sta99,sta01}. 
A possible explanation is that this problem is not so well
defined. Indeed, the concept of ``optimal'' $L_w$ is strictly linked to the
problem at hand. For example, if one is interested in tracking the ``long term''
evolution of a signal rather than possible short ``bursts'', then larger $L_w$'s
represent the ``best'' choice. The contrary is true if the short term evolution
is of interest. However, in certain applications (e.g. automatic signal
detection) it should be useful to have an ``objective'' method able to provide a
``good'' $L_w$ without the intervention of an operator. Of course, that requires
the adoption of an optimality criterion. At this regards, a possible choice is
that $L_w$ provides the representation with the ``best'' global time-frequency
resolution. In this case the following procedure is useful:
\begin{figure}
	\resizebox{\hsize}{!}{\includegraphics{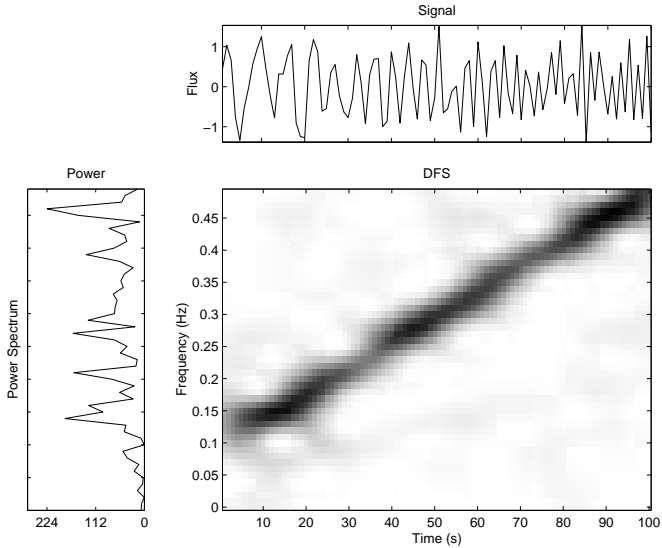}}
	\caption{ Grayscale image of the $\DFS[m,n]$ representation
	relative to the chirp signal $x(t) = \sin[2 \pi t \nu(t)]$ with $\nu(t) = 4.0
	\times 10^{-3} \times t + 0.10$ and $0 \leq t \leq 100$ sec., when sampled with a
	frequency rate of $1.0$ Hz and contaminated with a discrete, Gaussian white noise
	process with standard deviation equal to 0.3.}
	\label{fig:chirp}
\end{figure} 
\begin{itemize}
\item[-] a set, say $\Lambda$, of integer values for $L_w$ is fixed;
\item[-] the $\SDFS[m,n]$'s corresponding to the $L_w \in \Lambda$ are calculated;
\item[-] for each of these $\SDFS[m,n]$ the two-dimensional
       autocorrelation function $\ACF[k,l]$ is computed:
\begin{multline} \label{eq:ACF}
\text{$\ACF[k,l] =$ } \\
\shoveleft{\text{$= \sum_{m=0}^{M-1} \sum_{n=0}^{N-1} \SDFS[m,n] ~\SDFS[m+k,n+l]$}}
\end{multline}
with $-M < k < M$ and $-N < l < N$.
\item[-] the value of $L_w$ to which corresponds the ``sharpest''
          $\ACF[m,n]$ is selected as the searched length.
\end{itemize}
The idea behind such a procedure is that the concept of ``best'' global
resolution implies a representation maximally concentrated in the time-frequency
plane. Indeed, a value of $L_w$ too large or too small will results in a bad
resolution along the time or frequency dimension and therefore in a
representation that tends to spread in the time-frequency domain. The use of
$\ACF[k,l]$ stems by the observation that the larger the spread of the
representation, the more similar (correlated) the values are of the adjacent
elements in $\SDFS[m,n]$. This will reflect in $\ACF[k,l]$'s with significant
values also for lags $k, l$ different from zero.
\begin{figure}
	\resizebox{\hsize}{!}{\includegraphics{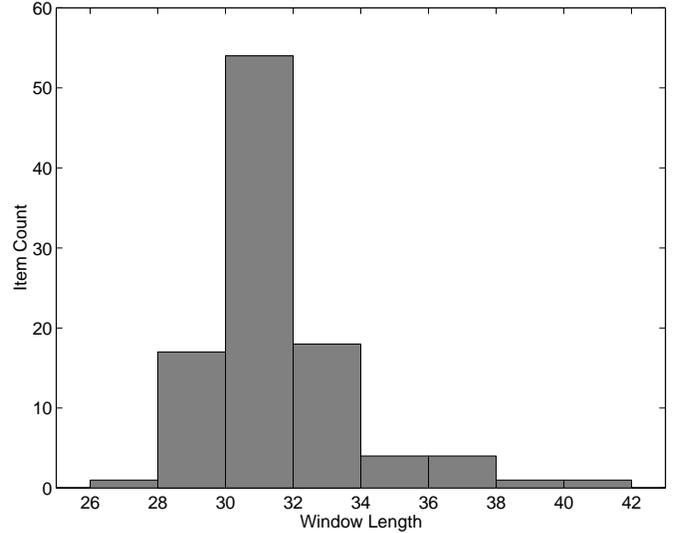}}
	\caption{ Histogram concerning the lengths $L_w$ provided
	in 100 simulations by the autocorrelation method described in the text. The
	signal used in this simulation it is shown in the uppermost panel of Fig.\ref{fig:chirp}
	and the target value is $L_w = 31$.}
	\label{fig:test_wind}
\end{figure} 
The only open question in this procedure is the measure of the ``sharpness'' of
$\ACF[k,l]$. At this regards there are various possibilities. In our numerical
experiments we have seen that the simple count of the elements of $\ACF[k,l]$
with values larger than a given threshold, say $50\%$ of the maximum, is able to
provide already satisfactory results.
This method for the estimation of $L_w$ works very well even in the situation of very
noisy data (see Figs.\ref{fig:chirp} and \ref{fig:test_wind}). The only concern
regards the calculation of $\ACF[k,l]$ that, if carried out on the basis of its
definition (\ref{eq:ACF}), can be computationally very expensive.
Indeed, for saving computing time, this operation has to be carried out in the Fourier domain
via the inverse double Fourier transform of the bidimensional power-spectrum of
$\SDFS[m,n]$. Although very fast, with this approach it is implicitly assumed
that $\SDFS[m,n]$ is periodic with regards to both indices; this fact introduces edge effects.
Fortunately, they can be avoided via zero padding; if interested in the correlation
for lags as large as $\pm K$ along both dimensions, then both ends of $\SDFS[m,n]$
must be zero padded in such a way to obtain a matrix with dimensions
$(M + K) \times (N + K)$.

\subsection{How to improve the resolution of the time-frequency representations?}

The main limitation of $\SDFS[m,n]$ is its bad resolution. In many practical
situations that is not a so serious problem. However, in certain applications 
(e.g. tracking of quasi-periodic sources) more precise
results can be required. This is a subject that has been extensively treated
in current literature. Here three methods are shortly presented that are able to 
provide interesting results. We warn, however, that this list is by no means complete 
and other important approaches are available in current literature as, for instance, 
the reduced interference distributions \citep{coh95}. 

\subsubsection{Wigner-Ville distribution}

\begin{figure}
	\resizebox{\hsize}{!}{\includegraphics{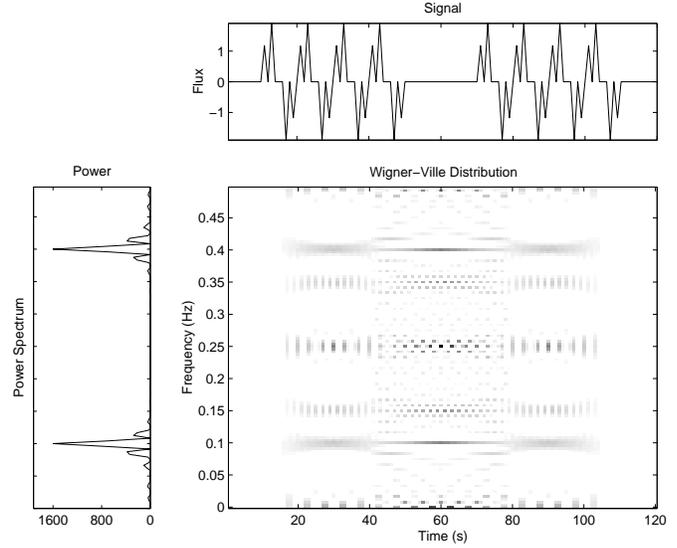}}
	\caption{ Grayscale image of the $\WVD[m,n]$ representation
	for the time series shown in the
	uppermost panel of Fig.\ref{fig:tf_dsinus2}.}
	\label{fig:tf_wv}
\end{figure}

The loss of resolution suffered by the $\FS(t,\nu)$ is due to the fact that, for
each time instant $t$, $\STFT(t,\nu)$ is calculated by using only a segment of
the signal. In principle, if such windowing were avoided, it could be possible
to obtain a JTFR with a much better resolution. Indeed such a JTFR does exist, 
it is named Wigner-Ville Distribution, and comes out from the problem of defining 
a representation, say $\WVD(t,\nu)$, with the property that:
\begin{eqnarray} \label{eq:prop1} 
\int\limits_{-\infty}^{+\infty} \WVD(t,\nu) ~d\nu &=& |x(t)|^2; \\
\int\limits_{-\infty}^{+\infty} \WVD(t,\nu) ~dt &=& \PS(\nu). \label{eq:prop2} 
\end{eqnarray}
In other words, the marginals with respect to frequency and time have to provide,
respectively, the squared signal and the corresponding power-spectrum. The form of this 
function, that however is not the only one sharing properties (\ref{eq:prop1}) and 
(\ref{eq:prop2}), is given by \citep{fla99, gro01}
\begin{multline} \label{eq:WVD}
\text{$\WVD(t,\nu) =$} \\
\shoveleft{\text{$=\int\limits_{-\infty}^{+\infty} x(t+\tau/2)~x^*(t - \tau/2)~
e^{-j 2 \pi\nu \tau}~d\tau.$}}
\end{multline}
In reality, the use of this JTFR has been rather limited because of two
drawbacks:
\begin{itemize}
\item[-] $\WVD(t,\nu)$ can go negative. In fact, except very particular cases, e.g., $x(t)$ 
given by a Gaussian function, it will go negative for
all signals. The non-positive property prohibits an
energy interpretation of this representation;
\item[-] the quadratic nature of $\WVD(t,\nu)$ determines the occurrence
of non-linear effects when dealing with multicomponents signals.
\end{itemize}
This last point is particularly troublesome. In fact, if $x(t)=g(t)+r(t)$, it happens that
\begin{multline}
\text{$\WVD_x(t,\nu) =$} \\
\shoveleft{\text{$=  \underbrace{\WVD_g(t,\nu) + \WVD_r(t,\nu)}_{\rm Auto-Terms} +
 2 \cdot \underbrace{Re \{ \WVD_{g,r}(t,\nu) \}}_{\rm Cross-Terms} $}}
\end{multline}
where
\begin{multline}
\text{$\WVD_{g,r}(t,\nu) =$} \\
\shoveleft{\text{$=\int\limits_{-\infty}^{+\infty} g(t+\tau/2) ~r^*(t - \tau/2) ~e^{-j
2 \pi \nu t} ~d\tau. $}}
\end{multline}
In general, the $\WVD(t,\nu)$ corresponding to a signal containing $N$ components will be 
characterized by $N$ {\it auto-terms}, that constitutes the desired information, 
and $0.5 \times N \times (N-1)$ real {\it cross-terms} which can be considered a sort of 
unwanted
interference disturb. The worst point is that the intensity of such {\it
cross-terms} is larger than those of the corresponding {\it auto-terms}. As shown
in Fig.\ref{fig:tf_wv}, that can mean unreadable results. In spite of these
problems, $\WVD(t,\nu)$ has very interesting properties not shared by most of the
other representations.
For this reason, in these last years, one of the most important subject in the field
of time-frequency analysis has been the research of methods able to reduce the
importance of the interference terms with the minimum deterioration of such properties.

\subsubsection{The reassignment method}

It is possible to show \citep{fla99} that the spectrogram $\FS(t,\nu)$ is the 2D-convolution 
of the Wigner-Ville distribution of the signal $x(t)$, say $\WVD_x(t,\nu)$ with the
Wigner-Ville distribution, say $\WVD_w(t,\nu)$ of the window function $w(t)$:
\begin{multline} \label{eq:SWVD}
\text{$\FS(t,\nu) =$} \\
\shoveleft{\text{$=\int\limits_{-\infty}^{+\infty} \int\limits_{-\infty}^{+\infty} 
\WVD_x(s,\xi)~\WVD_w(t-s,\nu - \xi) ~ds ~d\xi.$}}
\end{multline}
From this equation it is possible to see that the reason why $\FS(t,\nu)$, contrary to 
$\WVD(t,\nu)$, does not present the interference terms is due to the fact that such terms 
are smoothed out by the convolution. Such operation, however, deteriorates the 
time-frequency resolution.

\begin{figure}
	\resizebox{\hsize}{!}{\includegraphics{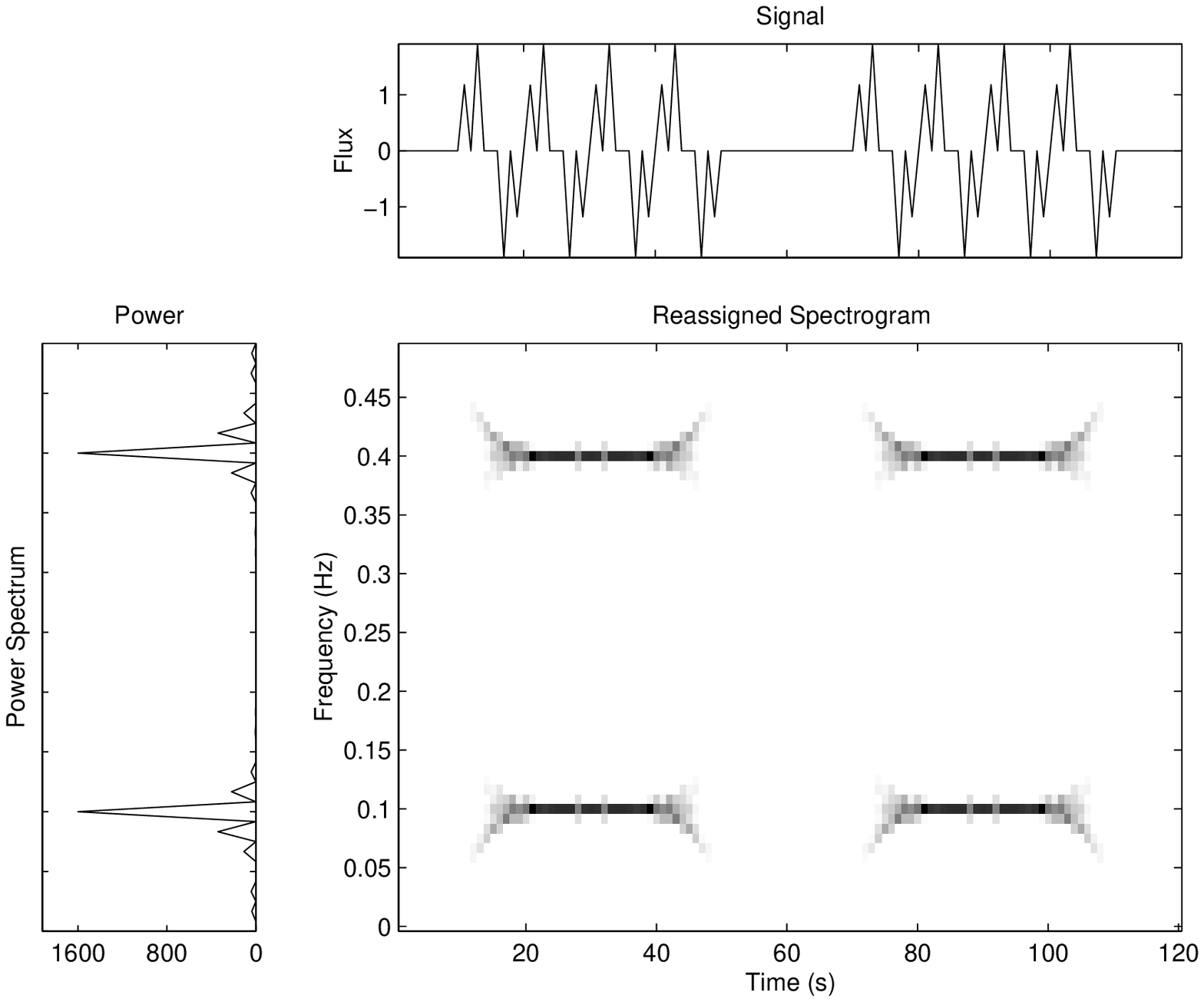}}
	\caption{ Grayscale image of the $\FSR[m,n]$ representation
	for the time series shown in the uppermost panel
	of Fig.\ref{fig:tf_dsinus2}.}	
	\label{fig:tfr_rs}
\end{figure}

A closer look at Eq.(\ref{eq:SWVD}) shows that $W_w(t-s,\nu-\xi)$ delimits a
time-frequency domain at the vicinity of the $(t,\nu)$ point, inside which a
weighted average of the $\WVD_x$ values is performed. This is the reason of the 
bad time-frequency resolution of $\FS(t,\nu)$. Indeed, the key point of the
reassignment principle is that these values have no reason to be symmetrical
distributed around $(t,\nu)$, which is the geometrical center of this domain;
their average should not be assigned at this point, but rather at the
center of gravity of this domain, which is much more representative of the local
energetic distribution of the signal. In practice, with the reassignment approach
one moves each value of the spectrogram computed at any point $(t,\nu)$ to
another point $({\hat t, \hat \nu})$ which is the center of gravity of the signal
energy distribution around $(t,\nu)$:
\begin{equation}
\hat t (t,\nu) = \frac { \displaystyle{ \int\limits_{-\infty}^{+\infty}
\int\limits_{-\infty}^{+\infty} s ~\WVD_w(t-s,\nu-\xi) ~\WVD_x(s,\xi) ~ds d\xi}} {
\displaystyle{ \int\limits_{-\infty}^{+\infty} \int\limits_{-\infty}^{+\infty}
\WVD_w(t-s,\nu-\xi) ~\WVD_x(s,\xi) ~ds d\xi}}
\end{equation}
\begin{equation}
\hat \nu (t,\nu) = \frac {\displaystyle{\int\limits_{-\infty}^{+\infty}
\int\limits_{-\infty}^{+\infty} \xi ~\WVD_w(t-s,\nu-\xi) ~\WVD_x(s,\xi) ~ds d\xi}}
{\displaystyle{\int\limits_{-\infty}^{+\infty} \int\limits_{-\infty}^{+\infty}
~\WVD_w(t-s,\nu-\xi) ~\WVD_x(s,\xi) ~ds d\xi}}
\end{equation}
and thus leads to a reassigned spectrogram, whose value at any point $(t',\nu')$
is the sum of all the spectrogram values reassigned to this point:
\begin{multline}
\text{$\FSR(t',\nu') =$} \\
\shoveleft{\text{$=\int\limits_{-\infty}^{+\infty} \int\limits_{-\infty}^{+\infty} \FS(t,\nu)
~\delta(t'- \hat t(t,\nu)) ~\delta(\nu'- \hat \nu(t,\nu)) ~dt ~d\nu $}} 
\end{multline}
Fig.\ref{fig:tfr_rs} show $\FSR[k,l]$ for the data shown in Fig.\ref{fig:sinus2}. A detailed 
description of the algorithm and of its digital implementation can be found in \citet{cha98}.

\subsubsection{Gabor spectrogram}

Recently, \citet{qia96} have presented an interesting technique for improving the 
time-frequency resolution of $\FS(t,\nu)$. They start from the consideration that each 
{\it cross-term} in $\WVD(t,\nu)$ is located midway the pair of corresponding 
{\it auto-terms} and tends to highly
oscillate in both time and frequency directions.
On the other hand, useful properties, such as the time marginal, frequency marginal
and instantaneous frequency, are obtained by averaging $\WVD(t,\nu)$.
For example, the mean
instantaneous frequency $\nu_{\rm inst}$, a quantity that is commonly used for
characterizing non-stationary signals, can be evaluated via:
\begin{equation}
\nu_{\rm inst} = \frac{\displaystyle \int \nu ~\WVD(t,\nu) ~d\nu}{\displaystyle
\int \WVD(t,\nu) ~d\nu}.
\end{equation}
These observations suggest that, if $\WVD(t,\tau)$ can be decomposed as the sum of
two-dimensional (time and frequency) localized harmonic functions, then we could
use the low-order harmonic terms to delineate the time-dependent spectrum with
reduced interference. High-order harmonics that introduce high oscillation have
relatively small averages and thereby have negligible influence on the useful
properties. The signal energy and useful properties are mainly determined by a
few low-order harmonic terms. On these bases, they propose the so called {\it
Gabor spectrogram} $\GS$ representation
\begin{equation}
\GS[k,l] = \sum_{d=0}^D P_d[k,l]
\end{equation}
where
\begin{multline}
\text{$ P_d[k,l] = 2 \sum_{A_d} a_{mn}~a^*_{mn} \times $} \\
\text{$ \exp \left\{ -\frac{\left[ k - \Delta M (m+m') /2 \right]^2}{2
\sigma_k^2}- \frac{\left[ l - \Delta N (n+n')/2 \right]^2}{2 \sigma_l^2} \right\} \times $} \\
\shoveleft{\text{$\exp \left\{ \frac{j 2 \pi}{L_w} \left[ k(n-n') \Delta N + \right. \right. 
$}} \\
\text{$\left. \left. + \left( l - \frac{n+n'}{2} \Delta N \right) \Delta M (m'-m) 
\right] \right\}, $}
\end{multline}
where
\begin{equation}
A_d = \{ (m,m'), (n,n') ; |m-m'| + |n-n'| = d \},
\end{equation}
\begin{equation}
\sigma_k = \frac{T_1}{2 \sqrt{\pi}}, \qquad \sigma_l = \frac{L_w}{T_1 2
\sqrt{\pi}},
\end{equation}
with $0 \leq m,m' \leq M_{tot} -1$, $-N/2 \leq n,n' \leq N/2 - 1$, $0 \leq k < L_s$,
and $ L_w/2 \leq l < L_w/2 -1$. Here, $T_1$ is a parameter whose value can be
chosen $\approx \sqrt{\Delta M \cdot N}$ and $a_{mn}$ are the Gabor coefficients
calculated via Eq.(\ref{eq:Loc_DFT1}) that here we rewrite in a slightly modified form
\begin{multline}
\text{$a_{mn} = e^{-j 2 \pi m n \Delta M /N} \times $} \\
\text{$ \sum_{k=0}^{L_w -1} w^*[k] ~x[k + m ~\Delta M] ~e^{-j 2 \pi k n /
N}, $}
\end{multline}
$0 \leq m \leq M_{T} = [(L_s - L_w)/\Delta M + 1]$, $0 \leq n < N$.
\noindent
It is possible to show that the larger the value of $D$ the better is the
resolution of the representation but, at the same time, the higher is the
contribution of the high-order harmonics in the decomposition of $\WVD(t,\nu)$.
That means that there is a trade-off between time-frequency resolution and the
importance of the interference terms. Usually $D=3$,$4$ is sufficient for many
practical applications. For the time series of Fig.\ref{fig:tf_dsinus2}, 
${\rm GS}_4[k,l]$ provides results very similar to those of Fig.\ref{fig:tfr_rs}.
A detailed description of the algorithm and of its digital implementation can be found in
\citet{mun97, mun01}.

\section{Time-variant filtering of discrete signals}

Although data exploration is an interesting application of time-frequency
analysis, the issue where such an approach shows all its potency is signal
filtering \citep{mat02, hla02}. Indeed, given the capability of a JTFR in characterizing a
non-stationary signal, we may expect that a filtering technique, developed within
this framework, be able to adapt its action to the {\it instantaneous} properties
of the signal itself. Since it is much easier to treat this subject in the context 
of the linear representations ($\WVD$, $\GS$ and $\FSR$ are bilinear representations 
carrying no phase information in a form easy to use), we shall concentrate in our 
discussion on the use of the Gabor transform.

\subsection{Some preliminary notes}

We have seen that the GR of a discrete signal corresponds to map a
one-dimensional array onto a two dimensional matrix.
Of course, in order to benefit of this kind of representation for signal
filtering purposes, it is necessary to be able to do the reverse job. It can be shown
that the relationship between a discrete signal $x[k]$ and the corresponding GR is given by
\citep{gro01}
\begin{equation} \label{eq:synthesis}
\displaystyle{x[k] = \sum_{m=0}^{M-1} \sum_{n=0}^{N-1} ~a_{mn} ~g[k - m \Delta M]
~e^{j 2 \pi k n / N}}
\end{equation}
with $0 \leq k < L_s$ and
\begin{equation} \label{eq:gabor}
\displaystyle{a_{mn} = \sum_{k=0}^{L_s-1} w^*[k - m \Delta M]~x[k]~e^{- j 2 \pi
k n / N}}
\end{equation}
with $0 \leq m < M,~0 \leq n <N$.
Here, the {\it synthesis function} $g[k]$ and the {\it analysis function} $w[k]$
have to satisfy the so called {\it biorthogonality condition}
(or {\it Wexler--Raz identity}).
\begin{equation}  \label{eq:wexler}
\displaystyle{\sum_{k=0}^{L_s - 1} w^*[k] ~g[k + p N] ~e^{-j 2 \pi k q / \Delta
M} = \frac{\Delta M}{N} ~\delta_p ~\delta_q}
\end{equation}
with $-\Delta N < q < \Delta N, ~ 0 \leq p < \Delta M$.
Pairs of functions satisfying Eq.(\ref{eq:wexler}) are termed {\it dual
functions} or windows. Since this formula is symmetric with
respect to $g[k]$ resp. $w[k]$, duality is a symmetric relation.

\begin{figure}
\resizebox{\hsize}{!}{\includegraphics{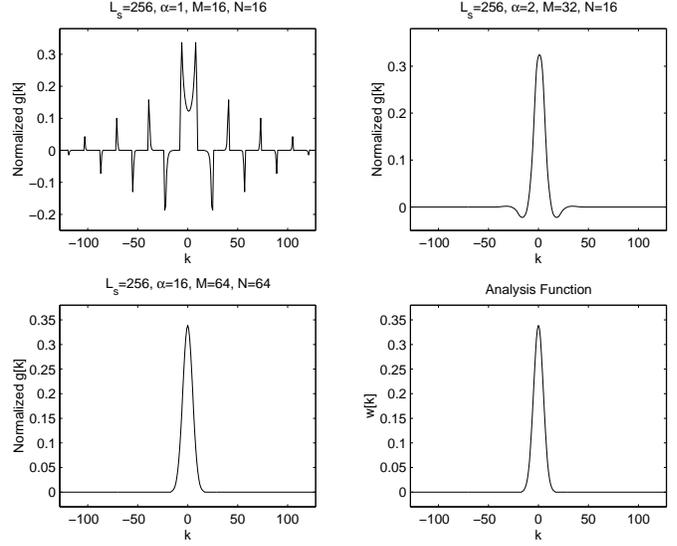}}
\caption{Gaussian analysis window $w[k]$ and the corresponding synthesis windows $g[k]$ 
for different values of the oversampling rate $\alpha$. For easyness of comparison, 
functions $g[k]$ are normalized to unit area.}
\label{fig:wind}
\end{figure}

For oversampled GR, with $\alpha > 2 L_w /(L_w + \Delta M)$, the system of linear
Eq.(\ref{eq:wexler}) is underdetermined. In other words, $g[k]$ and $w[k]$
do not determine each other in a unique way. Hence, for a given $g[k]$
there is a certain freedom in the choice of the corresponding dual complementary $w[k]$.
The solution suggested by \citet{qia96} is to solve system (\ref{eq:wexler}) with
the constraint, say $\Gamma$, that the shapes of $g[k]$ and $w[k]$ be as close as
possible (in least squares sense):
\begin{equation} \label{eq:gamma}
\Gamma = \min \sum_{k=0}^{L_s - 1}
\left (
\frac{w[k]}{\Vert w \Vert} -
\frac{g[k]}{\Vert g \Vert}  \right )^2.
\end{equation}
This choice turns out to correspond to the solution of the system (\ref{eq:wexler}) via a
{\it pseudo-inverse} method.
For example, by supposing $g[k]$ fixed and rewriting
system (\ref{eq:wexler}) in a matrix notation,
\begin{equation}
{\bf H} {\bf w}^* = {\bf u},
\end{equation}
the standard synthesis function ${\bf w}$, i.e., the window satisfying Eq.(\ref{eq:wexler}) 
with minimal norm, is given by
\begin{equation} \label{eq:dual}
{\bf w}^*={\bf H}^* ({\bf H} {\bf H}^*)^{-1} {\bf u}.
\end{equation}
It can be shown that the larger $\alpha$ the more similar are $g[k]$ and $w[k]$ 
(see Fig.\ref{fig:wind}). In particular, when $\alpha = L_s$ 
(i.e., $\Delta M = \Delta N =1$) then the two functions are identical. 

In case of a critical sampled GR (i.e. $\alpha = 1$,) $w[k]$ and $g[k]$ are univocally 
determined. However, according to the Balian-Low theorem, both the functions cannot be 
simultaneously well localized in time and frequency \citep[see, for example,][]{mal98}. 
In the practical application, the consequence is that $w[k]$ and $g[k]$ have very 
different shapes. This is the main reasons why in the filtering problems it is 
preferable to work with redundant representations (see below).

\subsection{Time--variant signal denoising} \label{sec:filter}

\begin{figure}
	\resizebox{\hsize}{!}{\includegraphics{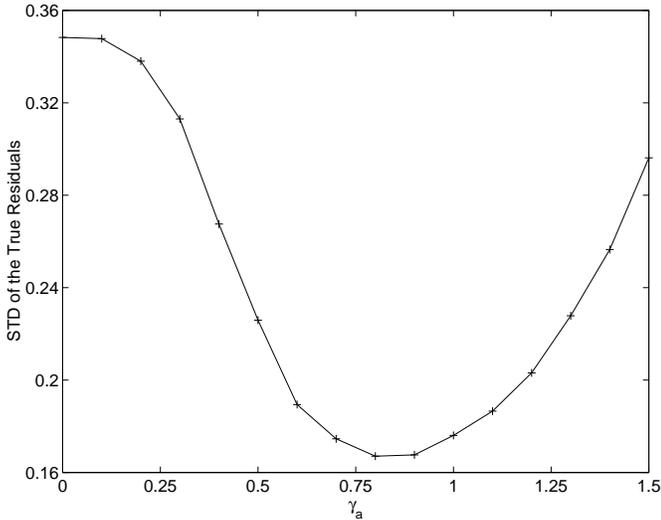}}
	\caption{ Dependence of the standard deviation (STD) of the
	residuals between the {\it true} and the reconstructed signals on the value of
	the noise threshold $\gamma_a$. The signal used in this example is that shown in
	the uppermost panel of Fig.\ref{fig:chirp}.}
	\label{fig:gamma_a}
\end{figure}

As shown in Fig.\ref{fig:gamma_a}, the choice of $\gamma_a$ is an important step
in a filtering procedure. Unfortunately, without substantial a priori
information (typical situation in many applications), there are no
so many ways to select the ``optimal'' value for this threshold.

In many practical situations the signal of interest is contaminated by a
broad--band random noise (white--noise). The key point is that such a noise tends
to spread evenly over the entire joint time-frequency domain, whereas the signal
component is usually concentrated in a relatively small region (e.g., see
Fig.\ref{fig:chirp}). This fact suggests a
simple procedure, similar to the hard thresholding method introduced by Donoho for wavelets 
\citep[see, for example, ][]{don95}, to filter out the noise contribution:
\begin{itemize}
\item[1)] determination of those indices for which the corresponding
   coefficients $a_{mn}$ may be considered as contributed by the noise.
   This step can be carried out in several ways according to the problem
   at hand. A typical  solution consists in setting a threshold value, say $\gamma_a^2$,
  for the elements $|a_{mn}|^2$  of $\SDFS[m,n]$. The idea is that the coefficients
  $a_{mn}$ of GR, whose absolute value is smaller than $\gamma_a$,
   can be assumed  unaffected by the signal contribution;
\item[2)] masking (zeroing) of such coefficients in order to obtain
   a ``cleaned'' version of the coefficients, denoted by $\tilde a_{mn}$,
   equal to zero in areas only occupied by noise;
\item[3)] synthesis of the filtered signal $\hat x[k]$ via Eq.(\ref{eq:synthesis}).
\end{itemize}
Although this approach is fairly simple, its application requires some care in
order to avoid unpleasant side effects.
In particular, one point is important:
\begin{quote}
once the analysis function $g[k]$ is fixed, it is advisable that the
synthesis function $w[k]$ be calculated via Eq.(\ref{eq:dual}).
In other words, the shapes of $g[k]$ and $w[k]$ have to be as close as possible.
\end{quote}
The good reason for this is that $g[k]$ is often chosen to have a good
localization both in time and frequency.
If $w[k]$ would not be forced to share this property it might be
badly localized along one or both of these two dimensions.
The consequence is that the synthesis of $\hat x[k]$, via the coefficient
$\tilde a_{mn}$, could provide different results from what expected;
\begin{figure}
	\resizebox{\hsize}{!}{\includegraphics{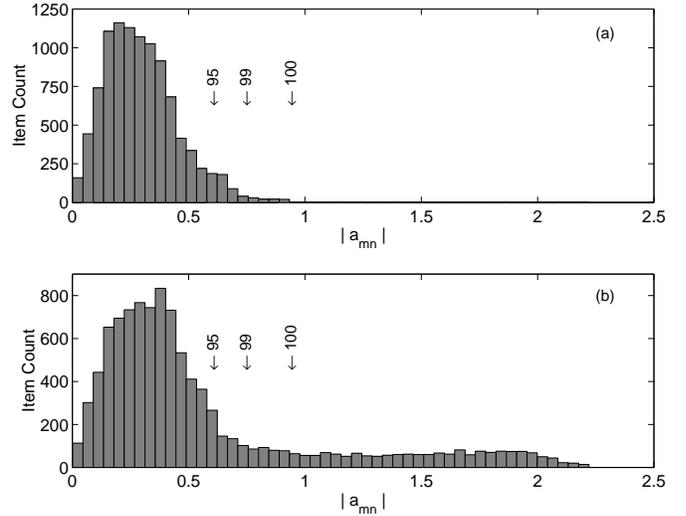}}
	\caption{ {\bf a)} Histogram of the $| a_{mn}
	|$ coefficients for a realization of a noise process with the same
	characteristics of that contaminating the signal of Fig.\ref{fig:chirp}. {\bf b)}
	Histogram of the $| a_{mn} |$ coefficients for the signal of
	Fig.\ref{fig:chirp}. The three arrows indicate the 95\%, 99\%, 100\% empirical
	percentiles relative to the histogram of the simulated noise process and provide
	three possible choices for $\gamma_a$.}
	\label{fig:thr_gamma_a}
\end{figure} 

Here it is necessary to stress that, in general, the set of modified coefficients $\tilde
a_{mn}$ will not constitute a valid {\rm GR} representation of any signal whatever. Indeed, 
it happens that
\begin{equation}
\hat x[k] \neq \sum_{m=0}^{M-1} \sum_{n=0}^{N-1} ~\tilde a_{mn} ~g[k - m \Delta
M] ~e^{j 2 \pi k n / N}.
\end{equation}
In particular the coefficients $\hat a_{mn}$, obtained from $\hat x[k]$ via
transform (\ref{eq:gabor}), can be different from zero even
for those indices
$m_0, n_0$, for which $\tilde a_{m_0 n_0} = 0$.
The reason is that coefficients $a_{mn}$ are obtained via a convolution operation 
and therefore no physical signal can correspond to a representation $\tilde a_{mn}$ 
characterized
by the sharp cutoffs created by the masking procedure.
In any case, it can be shown (see appendix \ref{sec:appendixa}) that,
if $w[k]$ and $g[k]$ satisfy the constraint (\ref{eq:gamma}),
$\hat x[k]$ is a signal whose Gabor
coefficients $\hat a_{mn}$ are as close as possible (in least squares  sense) to
$\tilde a_{mn}$.

A last note regards the fact that, in certain situations, the zeroing of coefficients $a_{mn}$
can constitute a too drastic operation. In fact, it can introduce some spurious
structures in the reconstructed signals. In this case, a possible alternative is 
represented by the weighting of coefficients $a_{mn}$, according to the expected level of noise 
contamination, in a way similar to that suggested by \citet{don95} (soft thresholding) in the 
context of wavelets.

\subsubsection{Practical determination of an ``optimal'' $\gamma_a$}

Apart the very common assumption of Gaussianity, the typical information
available about the noise is an estimate of its standard deviation. Conversely to
the case of classical Fourier analysis, the statistical characterization of a
pure random Gaussian process in the time-frequency domain is much more complex.
The problem is that coefficients $a_{mn}$ are not independent. This situation
suggests that the simplest (and safest) way to determine the threshold $\gamma_a$ is the
following:
\begin{itemize}
\item[-] simulation of a noise process (not necessarily Gaussian) via a pseudo-random 
generator;
\item[-] calculation of the corresponding $\SDFS[m,n]$ with the same modalities
used for the  calculation of the GR of the signal;
\item[-] estimation of a simple statistical index (e.g. empirical percentiles),
describing the characteristics of such a representation, that provides the
desired threshold for the $\SDFS[m,n]$ of the signal.
\end{itemize}
Although very simple, in our numerical simulations this method has proved to be
rather solid (see Fig.\ref{fig:thr_gamma_a} to compare with
Fig.\ref{fig:gamma_a}). An alternative to the last step could be the calculation
of a set of filtered time series $\hat x[k]$ corresponding to different choices
of $\gamma_a$. After that, among these solutions it is chosen the time series that
satisfies some statistical requirements as, for example, that the standard
deviation of the residuals between the original and the filtered signal is equal
to the nominal value of noise level. Although appealing, from our numerical
experiments this approach appears to be not so stable.

In the case one lacks also the information regarding the standard deviation of the
noise, the simplest solution consists in considering directly the $\SDFS[m,n]$ of
the signal and then in determining the threshold interactively by delimiting, via some 
graphical facility, the region where the signal is supposed to give its contribution. 
Another possibility is the use of some statistical indices of the values of $\SDFS[m,n]$ 
(e.g. percentiles).

\subsubsection{Could the SNR be further improved?}

The synthesis of $\hat x[k]$, via the coefficients $\tilde a_{mn}$, permits to
filter out the noise contribution corresponding to the coefficients $a_{mn}$ that
have been zeroed. However, what about the contamination of the remaining
coefficients? In their book, \citet{qia96} suggest a method that, according to
their opinion, should further improve the SNR of the final results. Their idea is
very simple and consists in an iterative application of the procedure described
in the previous section. In other words, once obtained the filtered $\hat x[k]$,
to this signal is applied the same procedure used for $x[k]$ and so on.
\begin{figure}
	\resizebox{\hsize}{!}{\includegraphics{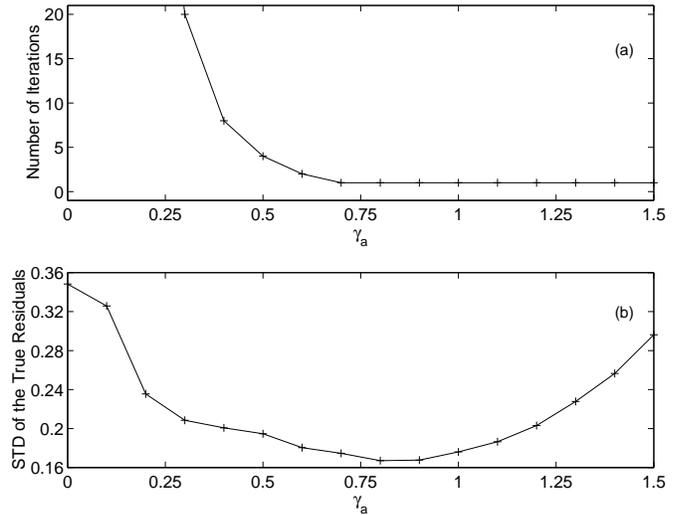}}
	\caption{ {\bf a)} Number of iteration for the Qian
	\& Chen method providing the best reconstruction of the signal of Fig.\ref{fig:chirp}
	as function of the noise threshold $\gamma_a$. {\bf b)} Standard deviation (STD) of the
 	corresponding residuals between the {\it true} and the reconstructed signals.}
	\label{fig:test_qc}
\end{figure} 

The most serious problem of this method consists in the lack of
theoretical argumentation that guarantee its reliability. Indeed, the repeated 
application of the masking procedure corresponds to a {\it contraction}, 
namely to an operation for which at
the {\it n}-th iteration
\begin{equation}
\| x_n[k] \|^2 < \| x_{n-1}[k] \|^2.
\end{equation}
In other words, the total energy of the filtered signal is continuously reduced
with the number of iterations. At first sight, one could accept this fact as
useful since effectively the noise tends to increase the power of an observed
time series. Unfortunately, such an increase regards the global characteristics
of a signal whereas the procedure suggested by Qian \& Chen is applied to the
coefficient $a_{mn}$ that reflect its {\it local} properties. In other words,
given an {\it observed} Gabor coefficient, the only thing one is allowed to claim
is that, because of the noise contamination, it will be different from the {\it
true} one. For a correct filtering, however, we should be able to decide at least
whether the contaminated coefficient is larger or smaller than the uncontaminated
one.

The only situation where the Qian \& Chen procedure can be expected to work
is when $\gamma_a$ is chosen too low. In situations like this one, a large
fraction of the unmasked coefficient $a_{mn}$ could be erroneously assumed linked
to the signal. In general, these coefficients are
smaller, in absolute value, than the coefficients effectively reflecting the signal
contribution. Consequently, the contraction connected with the Qian's \& Chen's
technique can be able to smooth out them before the signal component be too
negatively affected. However, we stress that, contrary to what claimed by the
two authors, the improvement is due to the removal of Gabor coefficients still
related to the noise rather that a better estimate of the coefficients
representing the contribution of the signal. Better results are to be expected
with a more appropriate choice of $\gamma_a$. All that is shown in the
Fig.\ref{fig:test_qc} to compare with Fig.\ref{fig:gamma_a}.

\subsection{Signal separation}
\begin{figure}
	\resizebox{\hsize}{!}{\includegraphics{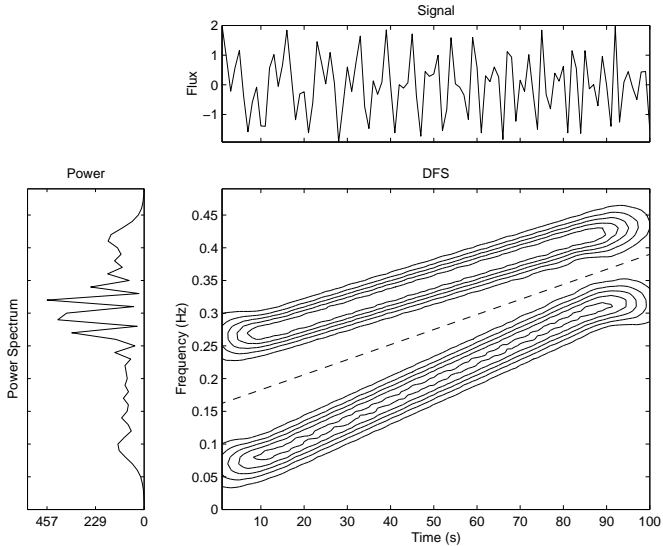}}
	\caption{ Contour plot of the $\DFS[m,n]$ representation
	for the signal mixture, given by two components disjoint in the time-frequency domain,
	described in the text. 
	Levels are uniformly distributed between the $10\%$ and $90\%$ of the peak value. 
	The dotted line delimits the two regions of the time-frequency domain that have been 
	used to separate the components shown in Fig.\ref{fig:test_chirp2}.}
	\label{fig:tf_chirp2}
\end{figure} 

\begin{figure}
	\resizebox{\hsize}{!}{\includegraphics{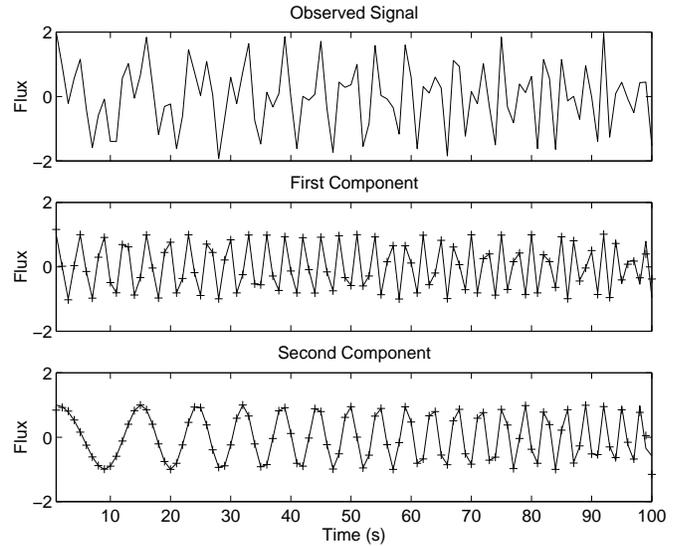}}
	\caption{ Observed signal, first component, and second 
	component corresponding to the $\DFS[m,n]$ representation of Fig.\ref{fig:tf_chirp2}. 
	Continuous line indicates the original signals whereas crosses indicate the 
	corresponding reconstructions obtained via the method described in the text.}
	\label{fig:test_chirp2}
\end{figure}

In certain practical situations it can happen that a given signal is the
mixture of two or more components. Often it can be of interest to separate the various 
contributions. In this respect, it can be shown \citep[][ and appendix 
\ref{sec:appendixb}]{hop00} that, within the context of linear filtering, 
this operation is possible (to a given degree of accuracy) only when a representation 
of signal is available where the various components are disjoint. For example, 
in the Fourier domain, two components are ``perfectly'' separable if their 
power-Spectra do not overlap. Therefore, in the context of the joint time-frequency 
representations, it can be expected that separability of signal mixtures is possible 
if the components do not overlap in the time-frequency domain (see Figs.\ref{fig:tf_chirp2} 
and \ref{fig:test_chirp2}). A simple and yet effective procedure is the same as that 
presented in section \ref{sec:filter}: each component is reconstructed by zeroing the 
Gabor coefficients corresponding to the remaining ones.

In order to understand the potential but also the limits of such an approach, it is 
necessary to understand what does it mean the fact that two signal are disjoint in the 
time-frequency domain. To discuss this problem,
let suppose to have a signal $x(t)$ given by the sum of two components,
say $x_1(t)$ and $x_2(t)$. If $\SDFS[m,n]$ is interpreted as an energy
distribution, no overlap in the time-frequency plane implies that
\begin{equation} \label{eq:energy}
\int | x(t) |^2 ~dt = \int |x_1(t)|^2 ~dt + \int |x_2(t)|^2 ~dt.
\end{equation}
In other words, the energy of the observed signal is equal to the sum of the
energies corresponding to the single components.
This point is very important since indicates that $x_1(t)$ and $x_2(t)$ can
be separated only if they do not interference and the total energy is conserved.
From the statistical point of view this condition means that the two components
are uncorrelated. In fact, if $x_1(t)$ and $x_2(t)$ are supposed, without loss
of generality, zero mean processes, then the integrals in Eq.(\ref{eq:energy})
provide the variance of the signals. However, it is well known that
\begin{equation}
\sigma^2_{x(t)} = \sigma^2_{x_1(t)} + \sigma^2_{x_2(t)} + 2 \sigma_{x_1(t)
x_2(t)}
\end{equation}
where the last term provides the covariance between $x_1(t)$ and $x_2(t)$.
Therefore, Eq.(\ref{eq:energy}) implies that $\sigma_{x_1(t) x_2(t)} = 0$.
\begin{figure}
	\resizebox{\hsize}{!}{\includegraphics{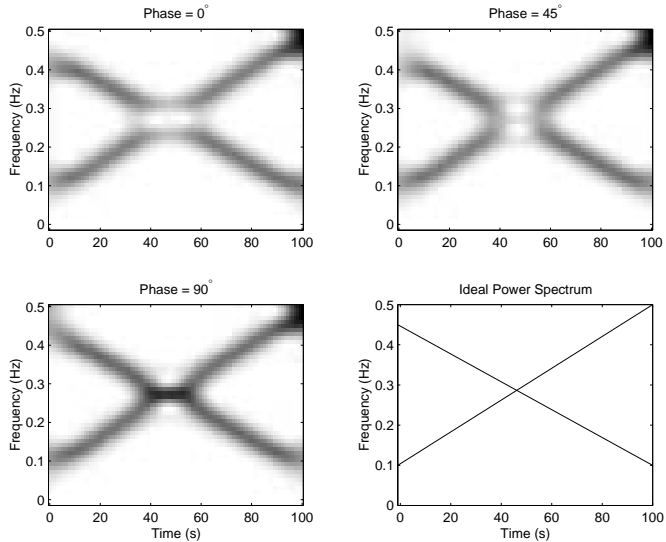}}
	\caption{ Grayscale image of the $\DFS[m,n]$ representation
	for three chirps described in the text and the corresponding ideal
	power-spectrum $\IPS(t,\nu)$.}
	\label{fig:phase}
\end{figure} 

From the above discussion it is evident that, in the separation of signals
overlapping in the time-frequency domain, things have to be expected much more
complex. For example, Figs.\ref{fig:phase} shows the $\SDFS[m,n]$ corresponding
to three signals, sampled with a frequency of one Hertz, that have been obtained
through the following models:
\begin{equation}
x(t) = \sin[2 \pi t \nu_a(t)] + \sin[2 \pi t \nu_b(t) + \phi_i]
\end{equation}
with $i=1,2,3$, $\nu_a(t) = 4.0 \times 10^{-3} \times t + 0.10$, $\nu_b(t) = -3.5 \times
10^{-3} \times t + 0.45$, $0 \leq t \leq 100$ sec., and $\phi_1 = 0$, 
$\phi_2 = \pi/2 $, and $\phi_3 = \pi$ {\rm rad}. 
In practice, the only difference lies in the phase $\phi$ of the second
component. The key point to note is that, although the $\SDFS[m,n]$'s are
different, these signals are characterized by the same $\IPS(t,\nu)$ (see
Fig.\ref{fig:phase}). In other words, in this specific case the equality
(\ref{eq:energy}) does not hold. Of course, that is the consequence of the
interference of the components present in $x(t)$. Unfortunately, the details of
such a interference will depend not only on the amplitudes of the components at a
given time instant, but also on the corresponding phases. Therefore, their
separation should require that each coefficient $a_{mn}$ be able to provide
information about four parameters. At least of some a priori information (e.g.
the value of the phases), this is clearly an underdetermined problem. The
situation becomes worst in case the number of overlapping signals is larger than
two.

One time more, these facts suggest that in order to obtain some insights on the
characteristics of a complex system, in general, the analysis of a signal, if not
developed within a given physical context, is absolutely insufficient.

\section{JTFR better than wavelets?}

\begin{figure}
	\resizebox{\hsize}{!}{\includegraphics{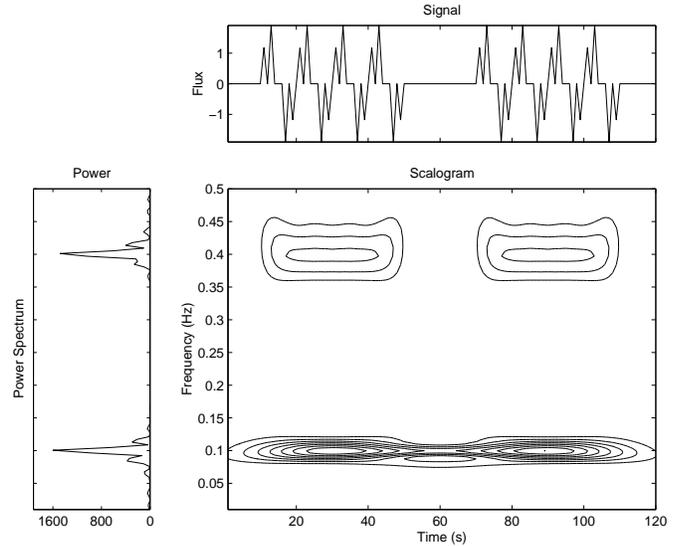}}
	\caption{Scalogram for the time series shown in the uppermost panel of 
	Fig.\ref{fig:tf_dsinus2}.}
	\label{fig:scalogram}
\end{figure}

Joint time-frequency analysis does not constitute the only possible tool for
the analysis of non-stationary signals.
Nowadays the so-called {\it Wavelet Analysis} arises as a new alternative to JTFA.
In the continuous case the ({\it Scalogram}) of a signal $x(t)$ is given by:
\begin{equation} \label{eq:scalogram}
\SCL(t,a)= | \WT(t,a) |^2
\end{equation}
with $W(t,a)$ being the continuous wavelet transform given by
\begin{equation} \label{eq:wavelet}
\WT(t,a) = \frac{1}{\sqrt{|a|}} \int\limits_{-\infty}^{+\infty} x(s) ~\Psi^* \left(
\frac{s-t}{a} \right)~ds.
\end{equation}
Here $\Psi(t)$ is a function, named {\it wavelet},
with the property that $\int \Psi(t) = 0$.
The idea behind such an approach is that, if $\Psi(t)$ is centered
in the neighborhood of $t=0$ and has a Fourier transform centered at a frequency
$\nu_0 \neq 0$, then in the time-frequency plane $\Psi \left[ (t-b)/a
\right]$ will be centered at $t=b$ and $\nu=\nu_0/a$, respectively. Consequently,
$\SCL(b,a)$ will reflect the signal behavior in the vicinity of such time instant
and frequency.
The main difference   with $\FS(t,\nu)$ is that the variable $a$
corresponds to a {\it scale} factor, in the sense that taking $|a| > 1$ dilates
$\Psi$ and taking $|a| < 1$ compress it. Therefore, in contrast to the analysis
function $w(t)$ in JTFR which maintains a fixed duration (or width) but changes
its shape due to frequency modulation, in wavelet analysis when the scale factor
$a$ is varied $\Psi(t,a)$ maintains its shape but changes its duration and
consequently its bandwidth. The practical consequence is that,
while $\FS(t,\nu)$ is characterized by a time-frequency resolution
constant over the entire time-frequency domain, $\SCL(t,a)$
presents a better time resolution
at high frequencies but a lower resolution at low frequencies 
(see Fig.\ref{fig:scalogram} to compare with Fig.\ref{fig:tf_dsinus2}).
At this point, the natural question arises concerning which choice, between the time-frequency
or wavelet approach, is the most appropriate one in the context of the analysis of 
astrophysical signals.
Of course, the answer strictly depends on the problem at hand.
In particular, it is useful to have a representation
characterized by a non-constant time-frequency resolution when the signal of
interest presents a non-stationary evolution on very different time scales as,
for example, in case of non-stationary self similar (fractal) signals.
On the other hand, especially in problems of exploratory data analysis, 
when the time behavior of a signal is dominated by
one or more non-stationary processes which evolve on similar time scales (e.g. a
periodic process that changes slowly its period), then it is by far better to use
a representation with a constant time-frequency resolution since it permits an
easier tracking of the changing characteristics of the signal. In the
astrophysical applications, very often this is the situation of interest.

It is necessary to mention that, especially in the field of {\it discrete}
wavelets, the above indicated limitations have been overcome in the sense that
some approaches have been devised that permit a large flexibility in the choice
of the resolution across the time-frequency plan \citep[e.g., wavelet packets, local
cosine bases, cf.][]{mal98}. In spite of that, in the exploratory analysis of the typical
astrophysical signals the joint-time frequency approach maintains its supremacy
since it is more intuitive and with better developed theoretical backgrounds
as far as the interpretation is concerned. Indeed, while JTFA essentially makes use of 
{\it localized} sinusoids, the shape of the
wavelets is much more difficult to explain from the physical point of view.
Consequently, the results provided by the time-frequency approach are more amenable 
for a direct physical interpretation.
\begin{figure}
	\resizebox{\hsize}{!}{\includegraphics{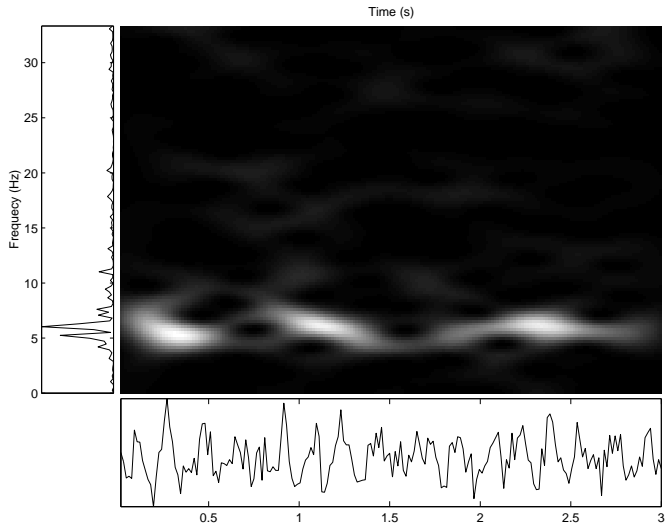}}
	\caption{Grayscale image of the DFS representation of a $3$ seconds segment extracted 	
	from the X-ray light curve of SCO-X1 described in the text.}
	\label{fig:scox1}
\end{figure} 

\section{Practical applications}

\begin{figure}
	\resizebox{\hsize}{!}{\includegraphics{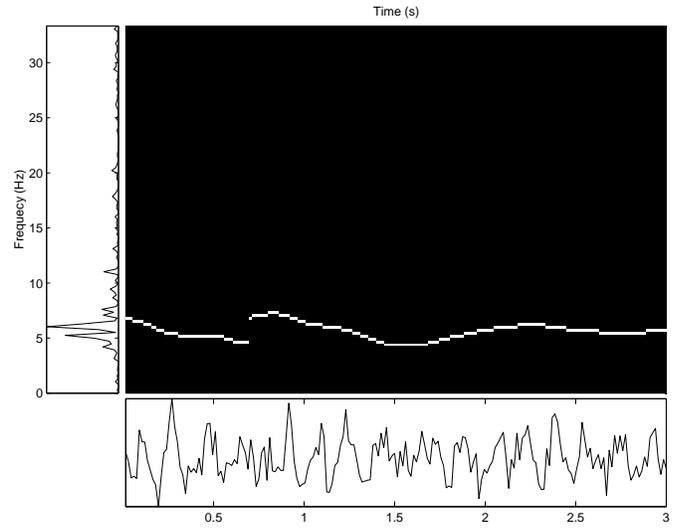}}
	\caption{Ridges of the grayscale image \ref{fig:scox1}.}
	\label{fig:scox1e}
\end{figure}

\begin{figure}
	\resizebox{\hsize}{!}{\includegraphics{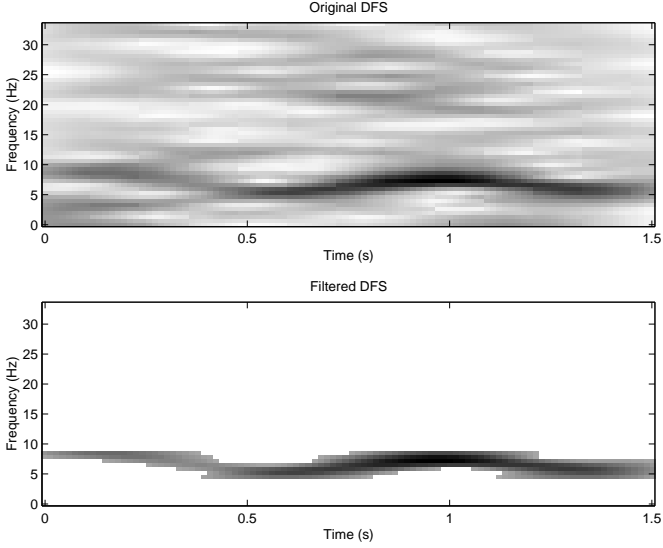}}
	\caption{Original vs. filtered DFS representation of a $1.5$ seconds segment 
	extracted from 	the X-ray light curve of SCO-X1 described in the text. Filtering 
	has been carried out via 	the method described in section \ref{sec:filter} with 
	$\gamma_a$ corresponding to the 	$95\%$ percentile of the DFS coefficients.}
	\label{fig:filt}
\end{figure}

\begin{figure}
	\resizebox{\hsize}{!}{\includegraphics{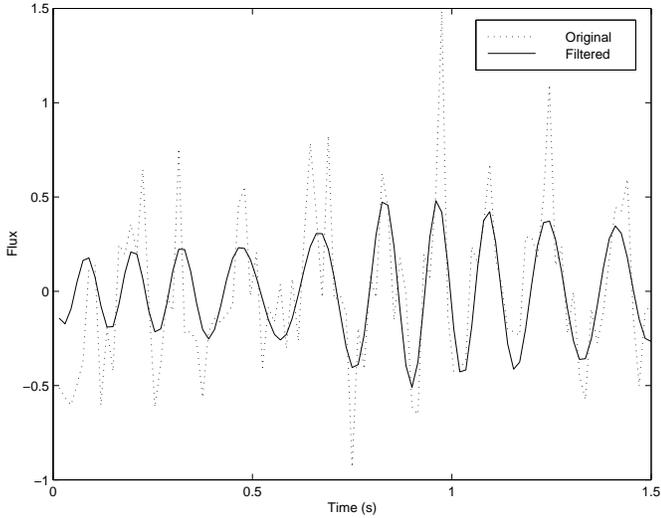}}
	\caption{QPO component vs. original time series obtained via the filtering shown in 
	Fig.\ref{fig:filt}}
	\label{fig:fserie}
\end{figure}

To illustrate the analysis strengths of JFTA in astronomical applications, we have applied 
some simple JFTA algorithms to a sequence of X-ray observations 
of low mass X-ray binary Sco X-1 (van der Klis, priv. comm.) made with the 
{\it Proportional Counter Array} (PCA) 
on the Rossi-XTE spacecraft \citep{bra93}. The time series used, shown 
in Fig.\ref{fig:scox1} (lower panel), is a 3 seconds subset of the available sequence that
is 1500 seconds long. Sampling is regular with a time step of 125 microsecond. As Sco X-1 
shows quasi-periodic oscillations (QPO) in the X-ray light emission,  
its light curve is characterized a periodic component that changes its 
period over time. Therefore, an important experimental activity consists in tracking 
such changes. Figs.\ref{fig:scox1},\ref{fig:scox1e} show that even a very simple JTFR as DFS 
is able to provide very interesting results, whereas Figs.\ref{fig:filt}, \ref{fig:fserie} 
show how it is possible to extract the component of interest from the observed signal. 
Although still preliminary and partial, this example represents
a good illustration of the possibilities of the application of the technique to 
astrophysical (and other) time series.

\section{Summary and conclusions}

In this paper we have considered the joint time-frequency domain as a possible framework 
for the analysis of the astrophysical signals. The methodologies based on such a domain 
appear very interesting since they allow reliable analyses and filterings of non-stationary 
time series that are very difficult, if not impossible, to carry out in the standard Fourier 
domain. These methodologies appear also useful for the separation of signal mixtures. 

\begin{acknowledgements}
We thank Prof. M. van der Klis for the availability of the unpublished data on Sco X-1 and 
Prof. H. Feicthinger for his useful comments.
\end{acknowledgements}

\appendix
\section{Masking of the Gabor Coefficients as Least Squares Filtering}
\label{sec:appendixa}

Transforms (\ref{eq:synthesis}) and (\ref{eq:gabor})
can be rewritten as matrix multiplications. In particular, the Gabor transform
({\ref{eq:gabor})
described as a linear mapping from the signal space to the coefficient space
${\cal C}^{MN}$ takes the form
\begin{equation} \label{eq:wmatrix}
{\bf a} = {\bf G^*_w} {\bf x}
\end{equation}
where ${\bf a} = {\rm vec} \{a_{mn} \}$ and ${\bf G_g}$ is an $(M N) \times
L_s$ matrix
\begin{equation}
{\bf G_w} = \left\{
\begin{array}{ccccccccccccccc}
\w \\ {\rm T}_{\Delta M} ~\w \\ {\rm T}_{2 \Delta M} ~\w \\ \vdots \\ {\rm T}_{M
\Delta M} ~\w \\ {\rm M}_{\Delta N} ~\w \\ {\rm M}_{\Delta N} {\rm T}_{\Delta M}
~\w \\ {\rm M}_{\Delta N} {\rm T}_{2 \Delta M} ~\w \\ \vdots \\ {\rm M}_{\Delta
N} {\rm T}_{M \Delta M} ~\w \\ {\rm M}_{2 \Delta N} ~g \\ {\rm M}_{ 2 \Delta N}
{\rm T}_{\Delta M} ~\w \\ \vdots \\ {\rm M}_{ N \Delta N} {\rm T}_{ (M-1) \Delta
M} ~\w \\ {\rm M}_{ N \Delta N} {\rm T}_{ M \Delta M} ~\w \\
\end{array}
\right\}
\end{equation}
where ${\rm M}_q$ and ${\rm T}_p$ are, respectively, the {\it modulation} and the
{\it translation} operators
\begin{equation}
\begin{array}{ll}
({\rm M}_q ~w)[k] &= w[k] ~e^{i 2 \pi q k /L_{s}} \\ ({\rm T}_p ~w)[k] &= w[k-p],
\end{array}
\end{equation}
and
\begin{equation}
({\rm M}_q {\rm T}_p ~w)[k] = w[k-p]~e^{i 2 \pi q k/L_s}.
\end{equation}
Since ${\bf G_w}$ is a full rank matrix, from Eq.(\ref{eq:wmatrix}) it easy to
show that the transform (\ref{eq:synthesis}) can be expressed in the form
\begin{equation} \label{eq:gmatrix}
{\bf x} = {\bf G}_{\bf w}^{\dagger} {\bf a},
\end{equation}
where
\begin{equation}
{\bf G}_{\bf w}^{\dagger} = ({\bf G}_{\bf w}^* {\bf G_w})^{-1} {\bf G}_{\bf w}^*
\end{equation}
is the {\it pseudo-inverse} matrix of ${\bf G_w}$ with dimensions $L_s \times (M N)$.

Here the key point is that, if in Eq.(\ref{eq:wmatrix}) we suppose ${\bf a}$ and
${\bf G_w}$ fixed and ${\bf x}$ unknown, then we are dealing with a system of $M
\times N$ equations in $L_s$ unknowns. For oversampled GR the number $M \times N$
is larger than $L_s$ and therefore the system of equations (\ref{eq:wmatrix})
is overdetermined. In this case, it can be shown that Eq.(\ref{eq:gmatrix}) 
delivers the corresponding least squares solution \citep{bjo90}.
The same still holds also when ${\bf a}$ is not in the
range of ${\bf G_w}$. That happens exactly, for example,
when the set of $a_{mn}$ does not constitutes a valid GR for $x[k]$.
In this case Eq.(\ref{eq:wmatrix}) has to be rewritten as
\begin{equation} \label{eq:ls}
{\bf \tilde a} \stackrel{LS}{=} {\bf G_w} {\bf \hat x}
\end{equation}
in order to highlight that the equality has to be intended in least squares
sense. However, $\hat x[k]$ will be characterized by a GR $\hat a_{mn}= {\bf G_w}
{\bf \hat x}$. Consequently, Eq.(\ref{eq:ls}) can be rewritten as
\begin{equation}
{\bf \tilde a} \stackrel{LS}{=} {\bf \hat a}.
\end{equation}
The meaning of this last expression is that, although $\tilde a_{mn}$ is not a
valid GR, $\hat x[k]$ synthesized via Eq.(\ref{eq:gmatrix}) corresponds to a
signal whose GR is as close as possible to $\tilde a_{mn}$ in least squares
sense.

\section{Signal Separability} \label{sec:appendixb}

In this section we briefly present the conditions according to which, given an observed 
signal $x(t) = d(t) + n(t)$, the component of interest $d(t)$ and the unwanted component 
$n(t)$ can be separated. 

A signal $x(t)$ may be represented on an arbitrary domain $\lambda$ by the following 
general integral transform: 
\begin{equation}
X(\lambda) = \int_T x(t) ~K(t,\lambda) ~dt \qquad \forall ~\lambda  \in \Lambda
\end{equation}
\begin{equation}
x(t) = \int_{\Lambda} X(\lambda) ~k(\lambda,t) ~d\lambda \qquad \forall ~t \in T
\end{equation}
where $\Lambda$ is the region in the signal space that the representation $X(\lambda)$ lies in
\footnote{In case of stochstic signals, integrals are to be interpreted in a mean-square 
limit.}. Here the functions $K(t,\lambda)$, called {\it direct transform kernel}, and 
$k(\lambda, t)$, called the {\it inverse transform kernel}, are linked by the relationships:
\begin{equation}
\int_{\Lambda} k(\lambda,t)~K(\tau,\lambda)~d\lambda = \delta(t-\tau)
\end{equation}
\begin{equation}
\int_{T} k(\lambda,t) ~K(t, \hat \lambda) ~dt = \delta(\lambda - \hat \lambda)
\end{equation}
For example in case of the Fourier transform $\lambda$ is equal to the Fourier frequency 
$\nu$, $K(t,\nu) = \exp(-i 2 \pi \nu t)$ and $k(\nu,t) = K^*(t,\nu)$.

If signal $d(t)$ has to be reconstructed by linear filtering of the observed signal $x(t)$ 
(i.e., by making use of linear operators), then it has to hold \citep[see, 
for example, ][]{hop00} 
\begin{equation} \label{eq:impulse}
\hat d(t) = \int_T h(t,\tau) ~x(\tau) ~d\tau,
\end{equation}
where $h(t,\tau)$ is the response at time $t$ given an impulse occurred at the filter input 
at time $\tau$. A sufficient condition for the existence of $h(t,\tau)$ is that it can be 
written in the form
\begin{equation}
\begin{array}{ll} \label{eq:condition}
h(t,\tau) &= \displaystyle{ \int_{\Lambda_d} k(\lambda,t) ~K(\tau,\lambda) ~d\lambda} \\
&+ \displaystyle{\int_{\Lambda_0} \int_{\Lambda_0} H_0(\lambda,\hat \lambda) ~k(\lambda,t) 
~K(\tau,\hat \lambda)~d\lambda ~d\hat\lambda}.
\end{array}
\end{equation}
Here $\Lambda_d$ is the region of the $\Lambda$ space over which the spectral component of 
$d(t)$ does not overlap the spectral component of $n(t)$, whereas $\Lambda_0$ is the spectral 
region where these two components overlap. $H_0(\lambda,\hat \lambda)$ is the solution of
\begin{equation} \label{eq:condition1}
P_{dx}(\lambda, \hat \lambda) = \int_{\Lambda_0} H_0(\lambda,\tilde \lambda) ~P_{xx}(\tilde 
\lambda, \hat \lambda) ~d\tilde\lambda, \qquad \forall~\lambda,\hat\lambda \in \Lambda_0
\end{equation}
where 
\begin{equation}
P_{yz}(\lambda,\hat\lambda) = \left\{
\begin{array}{ll} \label{eq:gpower}
Y(\lambda) Z^*(\hat \lambda) & \mbox{if $y(t)$, $z(t)$ deterministic}; \\
E[Y(\lambda) Z^*(\hat \lambda)] & \mbox{if $y(t)$, $z(t)$ stochastic}; 
\end{array}
\right.
\end{equation}
is the {\it generalized cross power-spectrum} concerning to the processes $y(t)$ and $z(t)$.
It can be shown \citep{hop00} that the reconstruction $\hat d(t)$, as provided by 
Eqs.(\ref{eq:impulse}) and (\ref{eq:condition}), is characterized by a variance 
$\sigma^2(t) = E[|\hat d(t) - d(t)|^2]$ equal to
\begin{equation} \label{eq:variance}
\sigma^2(t) = \int_{\Lambda_0} \int_{\Lambda_0} P_{\sigma \sigma}(\lambda,\hat\lambda) 
~k(\lambda,t) ~k^*(\hat\lambda,t) ~d\lambda ~d\hat\lambda
\end{equation}
with $P_{\sigma \sigma}$ the generalized power-spectrum of $\sigma^2(t)$.

Here the important points are three:
\begin{itemize}
\item[1)] the first term in the rhs of Eq.(\ref{eq:condition}) is independent of any 
signal. Moreover, it constitutes a so called {\it ideal filter} for $\Lambda_d$, namely a 
filter which passes without distorsion all {\it generalized frequencies} components falling 
in $\Lambda_d$ and rejects all others;
\item[2)] the second term in the rhs of Eq.(\ref{eq:condition}) depends, via 
Eq.(\ref{eq:condition1}), on the crosscorrelation function between $x(t)$ and $d(t)$ and 
therefore, since $x(t) = d(t) + n(t)$, on the crosscorrelation between $d(t)$ and $n(t)$;
\item[3)] the magnitude of $\sigma^2(t)$, given by Eq.(\ref{eq:variance}), depends 
on the extension of the overlap between $d(t)$ and $n(t)$. 
\end{itemize}
This last point is particularly interesting since it indicates that signals $d(t)$ and $n(t)$
can be perfectly separated only when they have disjoint supports in $\Lambda$. Indeed, 
in such a case, $\Lambda_0$ is an empty set and the three points above imply, respectively, 
that: 
\begin{itemize}
\item[-] Eq.(\ref{eq:condition}) simplifies in
\begin{displaymath}
h(t,\tau) = \displaystyle{ \int_{\Lambda_d} k(\lambda,t) ~K(\tau,\lambda) ~d\lambda};
\end{displaymath}
\item[-] $d(t)$ and $n(t)$ are uncorrelated signals;
\item[-] $\sigma^2(t) = 0$.
\end{itemize}
In other words, the perfect separation of two signals, via a linear filtering, requires the 
following condition be satisfied:
\begin{quote}
there exists some domain where the generalized spectral representation of the desired 
components are disjoint. Thus, the desired components must be uncorrelated processes.
\end{quote} 
All that holds also in the context of joint time-frequency analysis. The only different 
thing is that the function $h(t,\tau)$, with arguments in the time domain, has to be changed, 
via a Fourier transform with respect to $\tau$, in $H(t,\nu)$ which has arguments in the 
mixed time-frequency domain.


\begin{thebibliography}{}
\bibitem[Bradt et al. (1993)]{bra93} Bradt, H.V, Rotschild, R.E., \& 
Swank, J.H. 1993, A\&AS, 97,355
\bibitem[Bjorck (1990)]{bjo90} Bjorck, A. 1990, Numerical Methods for Least Squares 
Problems (SIAM, Philadelphia)
\bibitem[Chassande-Mottin (1998)]{cha98} Chassande-Mottin, E. 1998, Ph.D. thesis,
University of Cergy-Pointoise 
\bibitem[Cohen (1995)]{coh95} Cohen, L. 1995, Time-Frequency Analysis (Prentice Hall: London)
\bibitem[Daubechies (1992)]{dau92} Daubechies, I. 1992, Ten Lectures on Wavelets 
(SIAM: Philadelphia)
\bibitem[Donoho (1995)]{don95} Donoho, D.L. 1995, IEEE Transaction on Information Theory, 
41, 613
\bibitem[Duschl et al. (1991)]{dus91} Duschl, W.J, Wagner, S.J., \& Camenzind, 
M. 1991, Variability of Active Galaxies (Springer-Verlag: Berlin)
\bibitem[Feichtinger \& Strohmer (1998)]{fei98} Feichtinger, H.G., \& Strohmer, T.
1998, Gabor Analysis and Algorithms: Theory and Applications (Birkh\"auser: Boston)
\bibitem[Flandrin (1999)]{fla99} Flandrin, P. 1999, Time-Frequency/Time-Scale Analysis 
(Academic Press, London)
\bibitem[Gro\"chenig (2001)]{gro01} Gro\"chenig, K. 2001, Foundations of Time-Frequeny 
Analysis (Birkh\"auser: Boston)
\bibitem[Hlawatsch \& Matz (2002)]{hla02} Hlawatsch, F., \& Matz, G. 2002, in Time-Frequeny
Signal Analysis and Processing, ed. B. Boashash (Prentice Hall: London)
\bibitem[Hopgood (2000)]{hop00} Hopgood, J.R. 2000, Ph.D. thesis,
University of Cambridge 
\bibitem[Mallat (1998)]{mal98} Mallat, S. 1998, a Wavelet Tour of Signal Processing 
(Academic Press, London)
\bibitem[Matz \& Hlawatsch (2002)]{mat02} Matz, G., \& Hlawatsch, F. 2002, in Applications
in Time-Frequency Signal Processing, ed. A. Papandreu-Suppappola (CRC Press: Boca Raton)
\bibitem[Miller \& Wiita (1991)]{mil91} Miller, H.R., \& Wiita, P.J. 1991, 
Variability of Active Galactic Nuclei (Cambridge University Press: Cambridge)
\bibitem[Munk (1997)]{mun97} Munk, F. 1997, Ph.D. thesis, Aalborg University 
\bibitem[Munk (2001)]{mun01} Munk, F. 2001, compendium for the course 
``Joint Time Frequency Analysis'', SB8-2, Aalborg University 
\bibitem[Qian \& Chen (1996)]{qia96} Qian, S., \& Chen, D. 1996, 
Joint Time-Frequency Analysis (Prentice Hall: London)
\bibitem[Stankovic \& Katkovnik (1999)]{sta99} Stankovic, L., 
\& Katkovnik, V. 1999, IEEE Transactions of Signal Processing, 47, 1053
\bibitem[Stankovic (2001)]{sta01} Stankovic, L., 2001, Signal Processing, 81, 621
\bibitem[Vio et al. (1992)]{vio92} Vio, R., Cristiani, S., Lessi, O., 
\& Provenzale, A. 1992, ApJ, 391, 518 

\end{thebibliography}
\end{document}